\documentclass[12pt, letterpaper,aps,nofootinbib,floats,floatfix,amsmath,amssymb]{revtex4}
\usepackage{graphicx}
\usepackage{hyperref}
\usepackage{verbatim}
\usepackage{amssymb}
\usepackage{bm}

\begin{document}

\newcommand {\be} {\begin{equation}}
\newcommand {\ee} {\end{equation}}
\newcommand {\bea} {\begin{eqnarray}}
\newcommand {\eea} {\end{eqnarray}}
\newcommand {\eq} [1] {eq.\ (\ref{#1})}
\newcommand {\Eq} [1] {Eq.\ (\ref{#1})}
\newcommand {\eqs} [2] {eqs.\ (\ref{#1}) and (\ref{#2})}
\newcommand {\fig} [1] {fig.\ \ref{#1}}
\newcommand {\Fig} [1] {Fig.\ \ref{#1}}
\newcommand {\figs} [2] {figs.\ \ref{#1} and \ref{Halyo:1996pp#2}}

\def\lsim{\mbox{\raisebox{-.6ex}{~$\stackrel{<}{\sim}$~}}}
\def\gsim{\mbox{\raisebox{-.6ex}{~$\stackrel{>}{\sim}$~}}}

\def \eps {\epsilon}
\def \veps {\varepsilon}
\def \pl {\partial}
\def \hf {{1 \over 2}}
\def \mf {\mathbf}
\def\figdir{}
\def\sss{\scriptscriptstyle}

\renewcommand{\t}{\tilde}

\title{\bf Cosmic strings from pseudo-anomalous Fayet-Iliopoulos U(1)$_{\rm FI}$
  in D3/D7 brane inflation}

\author{Rhiannon Gwyn}
\email{rhiannon.gwyn@kcl.ac.uk}
\author{Mairi Sakellariadou}
\email{mairi.sakellariadou@kcl.ac.uk}

\author{Spyros Sypsas}
\email{spyridon.sypsas@kcl.ac.uk}
\affiliation{Department of Physics, King's College London,\\
Strand, London, U.K.  WC2R 2LS }

\date{\today}

\begin{abstract}
We examine the consequences of recent developments on Fayet-Iliopoulos
(FI) terms for D-term inflationary models. There is currently no known
way to couple constant FI terms to supergravity consistently; only
field-dependent FI terms are allowed. These are natural in string
theory and we argue that the FI term in D3/D7 inflation turns
out to be of this type, corresponding to a pseudo-anomalous U(1)$_{\rm
  FI}$. The anomaly is canceled by the Green-Schwarz (GS) mechanism in
4 dimensions. Inflation proceeds as usual, except that the scale is
set by the GS parameter $\delta_{\rm GS}$.  Cosmic strings resulting
from a pseudo-anomalous U(1) have potentially interesting
characteristics. Originally expected to be global, they turn out to be
local in the string theory context and can support currents. We outline
the nature of these strings, discuss bounds on their formation, and
summarize resulting cosmological consequences.
\end{abstract}
\maketitle



\section{Introduction}

The early universe, below the Planck energy scale, is described by an
effective ${\cal N}=1$ supergravity theory.  Since inflation should
have taken place at an energy scale $V^{1/4}\leq 4 \times 10^{16}$
GeV, any inflationary model should be built within supergravity. It is
however difficult to implement slow-roll inflation within
supergravity.  The positive false vacuum of the inflaton field
spontaneously breaks global supersymmetry, which is restored after the
end of inflation. In supergravity theories, the supersymmetry breaking
is transmitted to all fields by gravity, and thus any scalar field,
including the inflaton, acquires an effective mass of the order of the
expansion rate during inflation. This is the problem of Hubble-induced
mass (or $\eta$-problem), which originates from F-term interactions
and can be resolved if
the vacuum energy is dominated by the nonzero D-terms
of some superfields. This results in D-term inflation which, {\sl a
  priori}, can be easily implemented within string theory.

D-term inflation requires the existence of a nonzero Fayet-Iliopoulos
(FI) term, which can be added to the Lagrangian only in the presence
of an extra U(1) gauge symmetry. The symmetry breaking at the end of
the inflationary phase implies that cosmic strings (called D-term
strings) are always formed at the end of D-term hybrid inflation. By
tuning the free parameters of D-term inflation, the contribution of
cosmic strings to the Cosmic Microwave Background (CMB) temperature
anisotropies can be within the upper limit allowed by the
measurements~\cite{Rocher:2004my,Rocher:2006nh}.

Going from supersymmetric grand unified theories to string theory, one
can again build a D-term hybrid inflation model~\footnote{Note that in a 
realistic string model like the one studied here, one cannot avoid F-term 
contributions in the inflationary potential since these are needed to 
stabilize the moduli fields of the theory. These contributions can be 
fine tuned in order to avoid the $\eta$-problem~\cite{Haack:2008yb}.} 
and obtain at the end of inflation, realized through brane interactions, 
one-dimensional string-like objects (D-strings) analogous to the D-term ones.

In the context of string theory, brane interactions offer a plausible
framework for the realization of inflation.  Compactification to four
space-time dimensions leads to scalar fields and moduli which could
play the r\^ole of the inflaton field, provided they do not roll
quickly.  Brane inflation ends by a phase transition mediated by open
string tachyons. The annihilation of the branes releases the brane
tension energy that heats up the universe so that the hot big bang
epoch can take place.  Brane annihilations allow the survival of only
three-dimensional branes~\cite{Durrer:2005nz,Nelson:2008sv} with the
production of fundamental string-like objects, called cosmic
superstrings~\cite{Polchinski:2004hb,Davis:2005dd,
  Sakellariadou:2008ie,Sakellariadou:2009ev,Copeland:2009ga} in an
analogous way to cosmic strings, which are generically
formed~\cite{Jeannerot:2003qv} at the end of hybrid inflation in the
context of supersymmetric grand unified theories.

In what follows we study a particular D-term inflationary model, which
arises within the context of brane cosmology, and we then discuss some
of its phenomenological consequences. More precisely, we examine the
consequences of recent developments on FI terms in D3/D7 brane
inflation. In Sec.~\ref{FI-SUGRA} we summarize the current
understanding in building models with constant FI terms coupled to
gravity. In Sec.~\ref{FI-brane} we discuss the origin of
field-dependent FI terms in D3/D7 brane inflation. In
Sec.~\ref{CS-brane} we study cosmic strings formed at the end of this
brane inflation model under consideration. We analyze the nature of
strings formed at the end of D3/D7 brane inflation and then constrain
the free parameters from current observational data. We round up with
our conclusions in Sec.~\ref{conclusions}.

\section{FI terms in supergravity}
\label{FI-SUGRA}
The reasons for historical difficulties in attempting to couple models
with (constant) FI terms to gravity in a gauge-invariant way have
recently become clear~\cite{Komargodski:2009pc,
  Dienes:2009td}. Namely, in the presence of FI terms it is not
possible to construct a self-consistent Ferraro-Zumino (FZ)
multiplet~\cite{Ferrara:1974pz}, and consequently one cannot couple
the theory to supergravity in the usual ``old minimal" formalism. If
the theory possesses a continuous R-symmetry, the
R-multiplet~\cite{Komargodski:2010rb, Dienes:2009td} exists and can be used to couple
to supergravity according to the ``new minimal" formalism (see
Ref.~\cite{Komargodski:2010rb} and references therein). In both case the on-shell formulation possesses a continuous global symmetry. However,
consistent theories of quantum gravity do not have exact continuous
global symmetries.

Without alternative methods for coupling a theory to supergravity,
these observations amounted to a proof by contradiction that
consistent supergravity theories with constant FI terms were
impossible (see Ref.~\cite{Komargodski:2009pc, Dienes:2009td}). In fact, an alternative interpolating
supercurrent supermultiplet, the {\cal S}-multiplet, is defined even
when there is an FI term and no {\cal
  R}-symmetry~\cite{Komargodski:2010rb}. Gauging this leads to a
supergravity theory with an additional chiral superfield $\Phi$,
equivalent to that arrived at by first adding $\Phi$ to the original
theory such that the combined theory has an FZ multiplet and can be
gauged as usual~\cite{Komargodski:2010rb, Seiberg:2010qd}. Thus the
{\cal S}-multiplet allows one to avoid the problems sketched above by
rendering the FI term field-dependent. These are the only FI terms we
know how to include in a consistent supergravity theory at present. We
do not have a proof of inconsistency or a no-go theorem, but neither
do we know how to construct a consistent supergravity theory with
constant FI terms.

\section{FI terms in D3/D7 brane inflation}
\label{FI-brane}
Field-dependent FI terms are expected to arise naturally in string
theory~\cite{Dine:1987xk, Dine:1987gj, Atick:1987gy}. Field-dependent
FI terms can arise when the gauge group U(1)$_{\rm FI}$ develops an
anomaly upon compactification of the original string theory. Since
string theory is anomaly free, anomaly cancellation (via the GS
mechanism~\cite{Green:1984sg}) must take place in the lower
dimensional theory, and this gives rise to FI terms with $\xi \sim
\delta_{\rm GS} \sim {\rm Tr}\, Q$, where $Q$ is the charge operator
of the fields charged under U(1)$_{\rm FI}$~\cite{Atick:1987gy}. The
U(1) is then referred to as pseudo-anomalous.  To see that the FI term
in the D3/D7 inflationary model~\cite{Dasgupta:2002ew,
  Dasgupta:2004dw, Haack:2008yb} corresponds to this case, recall the
setup of this model, in which the distance between a D3- and a
D7-brane (in a Type IIB theory compactified on $K3 \times T^2/{\mathbb
  Z}_2$, with the D7-brane wrapping $K3$) plays the r\^ole of the
inflaton. The supersymmetry of the system is broken by nonselfdual
fluxes on the D7-brane, which give rise to an attractive potential
between the branes as follows: Consider a nonselfdual flux ${\cal F} =
F-B = dA-B $ (where $F_{mn}$ is the field strength of the vector field
$A_m$ and $B_{mn}$ is the pull-back of the NS-NS 2-form field) on the
D7-brane and in the directions perpendicular to the D3-brane. Writing
the flux components as~\cite{Dasgupta:2002ew}
\begin{eqnarray}
{\cal F}_{67} = \tan \, \theta_1\, \, \, ; \,\,\, {\cal F}_{89} = \tan
\theta_2~,
\end{eqnarray}
the attractive potential between the branes is 
\begin{eqnarray}
V &\sim& (\sin \theta_1- \sin \theta_2)^2 
\nonumber\\
&\sim& (\theta_1 - \theta_2)^2~,
\end{eqnarray}
corresponding to the square of the supersymmetry breaking parameter
$\xi =\theta_1 - \theta_2$. The strings stretching between the D3-
and D7-branes in the $4,5$ directions correspond to the
waterfall fields (in the ${\cal N} = 2$
hyper-multiplet~\cite{Dasgupta:2002ew, Dasgupta:2004dw}), which are charged under
U(1)$_{\rm FI}$. We argue that the sum of their charges is $\xi =
\theta_1 - \theta_2$, by T-dualising along the $6,8$ directions to get
a D5-brane tilted by an angle $\theta_1 = \tan^{-1} (2 \pi \alpha'
{\cal F}_{76} )$ in the $(6,7)$-plane and by an angle $\theta_2 =
\tan^{-1} (2 \pi \alpha' {\cal F}_{89})$ in the $(8,9)$-plane. When the
angles are not equal, supersymmetry is broken in the T-dual picture as
it is in D3/D7~\cite{Berkooz:1996km, Herdeiro:2001zb}; this
corresponds to a rotation of the complexified scalars $\zeta_1 = x^6 +
\imath x^7$ and $\zeta_2 = x^8 + \imath x^9,$ according to $\zeta_1
\rightarrow e^{\imath \theta_1 } \zeta_1$ and $\zeta_2 \rightarrow
e^{\imath \theta_2} \zeta_2$, respectively.  Thus we can identify the angles
$\theta_i$ with the charges of the waterfall fields $\zeta_i$ under
the relevant U(1) in D3/D7.

The connection between nonselfdual flux and a pseudo-anomalous U(1)
was alluded to in Refs.~\cite{Binetruy:2004hh, Burgess:2003ic}, where
it was pointed out that the r\^ole of the axion in the GS mechanism in
Type IIB string theory will be played by the field dual to the 4-form
$C_{(4)}$, in the same multiplet as the volume modulus. Thus the
r\^ole usually played by the axion-dilaton is played in D3/D7, which
we analyze below~\footnote{Here $C_{(4)}$ denotes the Hodge dual of
  $B_{\mu \nu}$ where $C_{\mu \nu a b} = B_{\mu \nu} J_{ab}$, with
  $J_{ab}$ the K\"ahler form of $K3$ and $\mu, \nu$ denoting indices in
  the noncompact directions.}, by the complex K\"ahler modulus $s =
{\rm Vol}(K3) + \imath C_{(4)}$. The original construction using
nonselfdual fluxes on the D7-brane to generate an FI term was
motivated by the need to create a nonprimitive flux, to break
supersymmetry and to get an attractive potential between the
branes. We have seen that this construction is also intrinsically
consistent with having a field-dependent FI term, as demanded by our
current understanding of FI terms in supergravity.

\section{Cosmic strings in D3/D7 brane inflation}
\label{CS-brane}

The main implication of a pseudo-anomalous U(1)$_{\rm FI}$ in D3/D7
brane inflation is that the strings which form at the end of inflation
will be local axionic strings. Note that cosmic strings from
pseudo-anomalous U(1) might be naively expected to be global because
they are axionic; they are sourced by a 2-form which is Hodge dual to
a scalar $a$ with axionic 4-dimensional coupling~\cite{Casas:1988pa,
  Harvey:1988in}. However, the GS counter terms also give rise to a
coupling $\partial_\mu a A^\mu$ between the axion-type field and the
gauge field associated with the anomaly, as well as a mass term for
the gauge field~\cite{Dine:1987xk}. This corresponds to a higgsing of
the axionic instability~\cite{Copeland:2003bj}, rendering the axionic
strings local (with finite energy per unit
length)~\cite{Binetruy:1998mn, Davis:2005jf}. These local axionic
strings were studied further in Ref.~\cite{Davis:2005jf}, where the
conditions under which they will be current carrying were
discussed. We apply this analysis to the D3/D7 brane inflation model,
and then discuss the micro-physical nature of the strings formed at
the end of D3/D7 inflation, when the separation between the branes
goes subcritical and the strings stretching between them become
tachyonic. This corresponds in the gauge theory to the waterfall stage
at the end of inflation and ends when the D3- dissolves into the
D7-brane.

\subsection{String construction}
The D3/D7 brane setup preserves ${\cal N} = 2$ supersymmetry, and
inflation in this system is naturally described in ${\cal N} = 2$
language (i.e. as a specific case of P-term inflation
\cite{Kallosh:2001tm}). Equivalently, it can be described in ${\cal N}
= 2$ written in terms of ${\cal N} =1$ variables~\footnote{In fact the
  ${\cal N} = 2$ supersymmetry of the system is broken to ${\cal N}
  =1$ by bulk 3-form fluxes, while the remaining ${\cal N} =1$
  supersymmetry is broken by the worldvolume fluxes which give rise to
  the Fayet-Iliopoulos term.}~\cite{Dasgupta:2002ew, Haack:2008yb}. Following Ref.~\cite{Davis:2005jf}, we can therefore write the
relevant part of the ${\cal N }= 1$ action for the case of a
pseudo-anomalous U(1)$_{\rm FI}$ as
\begin{equation}
{\cal L}  =  Z_i^\dag e^{2 q_i V} Z_i + {\cal K}  +
\left [\frac{1}{4} S W^\alpha W_\alpha + W(Z_i, S) \right ]\delta
(\bar \theta^2) + {\rm h.c.}~,
\end{equation}
so that the bosonic part reads
\begin{eqnarray}
\label{BDDaction}
\nonumber{\cal L}_{\rm B} & = & |{\cal D}_\mu
\zeta_i|^2 - F_i^\dag F_i + \frac{1}{4 s_R^2} (\partial_\mu s_R)^2 +
\frac{1}{4 s_R^2} ( \partial_\mu a - 2 \delta_{\rm GS} A_\mu)^2 -
\frac{1}{4s_R^2} |F_s|^2 \\&&- \frac{s_R}{2} D^2 - \frac{s_R}{4}
F_{\mu \nu} F^{\mu \nu} +\frac{a}{4} F_{\mu \nu} \tilde F^{\mu \nu},
\end{eqnarray}
where $S(s, 2 s_R \chi_\alpha, F_s)$ is the chiral field associated
with the GS mechanism, $V(A_\mu, s_R^{- \frac{1}{2}} \lambda_\alpha,
D)$ is the U(1)$_{\rm FI}$ vector superfield and $Z_i(\zeta_i, \psi_{i
  \alpha}, F_i)$ are the chiral superfields that contain the waterfall
scalars. The covariant derivative is given by ${\cal D}_\mu \phi =
(\partial_\mu + \imath q_\phi A_\mu) \phi$. The $\zeta_i$ are the
fields charged under U(1)$_{\rm FI}$; these are known as waterfall
fields and appear in an ${\cal N} = 2$ hypermultiplet. In the D3/D7
setup they are given by the strings stretching between the D3- and
D7-branes~\cite{Dasgupta:2004dw}. By the argument given in
Sec.~\ref{FI-brane}, we can take their charges to be $q_1 = \theta_1$
and $q_2 = - \theta_2$. Further, as we argued above, $s = {\rm
  Vol}(K3) + \imath C_{(4)}$, i.e. the K\"ahler modulus~\footnote{We
  use the notation of Ref.~\cite{Haack:2008yb} multiplied by a factor
  of $-\imath$ in order to be consistent with the notation of
  Ref.~\cite{Davis:2005jf}.} of $K3$ will play the r\^ole usually
played by the axion-dilaton (see e.g. Ref.~\cite{Davis:2005jf}).
  
Here ${\cal K}$ represents an expansion in terms of derivatives of the
modified K\"ahler potential $K$~\cite{Buchbinder:1998qv}, which is given
by~\cite{Witten:1985xb}
\begin{eqnarray}
K(S, \bar S) & = & - \log\Big(S + \bar S - 4 \delta_{\rm GS} V + (Y_3
- \bar{Y}_3)^2\Big)~,
\end{eqnarray}
where we have included the dependence on the inflaton
$y_3$~\cite{Haack:2008yb} ($Y_3$ is the corresponding superfield) and
the modification due to the GS anomaly cancellation~\cite{Dine:1987xk,
  Binetruy:1998mn}; $\delta_{\rm GS}= (1/192 \pi^2) \sum q_i$. 

Note that the shift symmetry of the inflaton displayed by the K\"ahler
potential can be spoilt by quantum corrections. As discussed in
Ref.~\cite{Haack:2008yb}, some fine-tuning of the other moduli is then
required in order for the real part of $y_3$ to be a well-defined
inflaton. Of more relevance here is the fact that, via the dependence
of the D7-brane coupling on $y_3$, the appropriate definition of
$Re(s)$ also involves a term dependent on $y_3$. This will not affect
our argument because we do not make use of the specific form of $s$,
and we set $y_3=0$ in constructing our string solutions below.

The superpotential $W$ is of the form~\cite{Haack:2008yb}
\begin{equation}
W(y_3,\zeta_i,s,t)=W_0+A(\zeta_i,t,...)
\Big[1-\Delta(t)y_3^2+\mathcal{O}(y_3^4)\Big]e^{-ics},
\end{equation}
where $W_0$ is the flux superpotential, $A$ depends in general on all
moduli fields in the theory, such as the complex structure modulus $t$
of $T^2$, and the axion-dilaton $u$. Note that in our case $s_R$ {\sl
  does not} correspond to the dilaton, as in Ref.~\cite{Davis:2005jf},
but to the volume of $K3$, which the nonperturbative superpotential is
needed to stabilize. As emphasized in
Refs.~\cite{Binetruy:2004hh,ArkaniHamed:1998nu}, the volume modulus or
dilaton must be carefully stabilized in models with a field-dependent
FI term, so as not to disrupt the mechanism of D-term inflation. In
the D3/D7 case, the volume modulus is stabilized by a nonperturbative
superpotential due to gaugino condensation on a stack of D7-branes
wrapping the K3~\cite{Haack:2008yb}. The resulting F-term potential,
\begin{equation}
V_F = \frac{|A|^2 e^{-2 a s_R}s_R}{u_R} \left [\frac{3 a^2}{2 t_R} +
  F(y_3, t) \right ]~,
\end{equation}
where $F(y_3,t)$ is an even polynomial of the inflaton with
$t$-dependent coefficients~\cite{Haack:2008yb} and the subscript $R$
denotes the real part of the relevant field, can then stabilize the modulus $s$. By
taking into account stringy corrections, it was shown in
Ref.~\cite{Haack:2008yb} that this F-term potential will not disrupt
the flatness of the inflation potential.

The D-term is~\cite{Haack:2008yb}
\begin{eqnarray}
D & = & -\frac{1}{s_R} \left (q_i \zeta_i^\dag \zeta_i +
\frac{\delta_{\rm GS}}{s_R}\right )\nonumber\\ & = & - \frac{1}{s_R}\left
(\theta_1|\zeta_1|^2-\theta_2|\zeta_2|^2+ \xi \right )~,
\end{eqnarray}
where $s_R = {\rm Vol}(K3) = 1/g^2$, with $g$ the gauge coupling. Note
that $\xi$ and $g$ are the same as in Refs.~\cite{Binetruy:1996xj,
  Haack:2008yb}, and are related to the (tilded) values in
Refs.~\cite{Kallosh:2001tm, Dasgupta:2002ew, Dasgupta:2004dw} by $\xi
= (2 \tilde \xi)/\tilde g$ and $g = (1/2)\tilde g$. The D-term is
minimized by
\begin{equation}
|\zeta_1|=0,\quad|\zeta_2|=\eta,\quad s_r\equiv {{\rm
    Vol}(K3)}=\frac{\delta_{\rm GS}}{\theta_2\eta^2}~,
\end{equation}
where $\xi=g^2\delta_{\rm GS}=(\theta_1-\theta_2)/(192\pi^2)g^2$.

We take as our string {\sl ansatz}
\begin{eqnarray}\label{stringansatz}
 \zeta_1&=&0\nonumber\\ \zeta_2&=&\eta f(r)e^{\imath
   n\varphi}\nonumber\\ 
s&=&\frac{\delta_{GS}}{\theta_2\eta^2\gamma(r)^2}+2in\delta_{\rm
   GS}\varphi\nonumber\\ A_\varphi&=&n\frac{u(r)}{r}~,
\end{eqnarray}
where $f(r), \gamma(r), v(r) \rightarrow 1$ as $ r \rightarrow
\infty$, and $f(0) = \gamma(0) = v(0) =0$ (we take $y_3 = 0$). As in Ref.~\cite{Davis:2005jf}
the strings are coaxial, with an inner core of radius $r_{\rm D} \sim
m_{\rm D}^{-1}$ and an outer core of $r_{\rm F} \sim m_{\rm F}^{-1}$, where
\begin{equation}
m_{\rm D}=\sqrt{2\delta_{\rm GS}}g\theta_2 \sqrt{1+\theta_2^2}~,
\end{equation}
and
\begin{equation}
m_{\rm F} =  \sqrt{2} |A| e^{- a <s_R>} \sqrt{12 a^3 (a<s_R> - 1)}~,
\end{equation}
coming from the D-term and F-term potentials, respectively.

\subsection{Nature of the strings}
To better understand the nature of strings formed at the end of D3/D7
brane inflation, we examine their conduction properties. The fermionic
zero modes ($E=0$) can be found by performing a supersymmetry
transformation (with transformation parameter $\kappa_a$) on the
string solution, Eq.~(\ref{stringansatz}), since this leaves the
energy invariant~\cite{Davis:1997bs}, giving
\begin{eqnarray}
\delta \Psi_{1\alpha} & = & \sqrt{2} F_1 \kappa_{\alpha}\\ \delta
\Psi_{2 \alpha} & = & \sqrt{2} F_2 \kappa_{\alpha} + \imath \sqrt{2}
\kappa^\star_{\alpha} e^{ \imath(n \mp 1) \varphi} \eta \left [ f'(r)
\pm \frac{nf}{r} (1 - \theta_2 u(r) )\right]\\ \delta \lambda_\alpha
& = & \sqrt{\frac{\delta_{GS}}{\theta_2 \eta^2 \gamma^2}}\imath
\kappa_\alpha \left [ \frac{\theta_2^2 \eta^4 \gamma^2}{\delta_{GS}}
  (f^2 - \gamma^2) \mp \frac{n}{r} \left(u' - \frac{1}{r^2}\right)\right]
\\ \delta \chi_\alpha & = & \frac{1}{\sqrt{2}} \frac{\theta_2 \eta^2
  \gamma^2}{\delta_{GS}} \left [ F_s \kappa_\alpha - \imath 2
  \delta_{GS}\kappa^\star_\alpha e^{\mp \imath \varphi} \left
  (\frac{\gamma'}{\theta_2 \eta^2 \gamma^2} \mp \frac{n(1-u)}{r}
  \right ) \right ] \\ \delta \Psi_{3 \alpha} & = & \frac{1}{\sqrt{2}}
\frac{\theta_2 \eta^2 \gamma^2}{\delta_{GS}} F_3 \kappa_{\alpha}~,
\end{eqnarray}
where the upper (lower) sign corresponds to $\alpha =1 (2)$ and $F^i =
\partial \bar W/\partial \bar z_i$. We see that when the F-terms are
nonzero, as is the case for a general superpotential, these strings
break supersymmetry in the core, despite being local in the sense of
having finite energy~\cite{Davis:2005jf}. Their coaxial nature
complicates things somewhat: in the inner core of radius $r_{\rm D}$
the F-terms are zero and some supersymmetry is preserved, as is usual
for D-term strings. Only one zero mode is normalizable, meaning that
the strings are chiral, as in Ref.~\cite{Davis:2005jf}. This chirality
is lost when $\delta_{\rm GS} \rightarrow 0$, as expected for F-term
strings~\cite{Jeannerot:2004bt}. However, it is expected that at least
some zero modes will be higgsed by the gravitino transformation when
the theory is gauged~\cite{Jeannerot:2004bt, Brax:2006yb}. In this
case current can only arise from suitable Yukawa couplings, which seem
to be generic in the context of our analysis.

Chiral local axionic strings could have interesting properties, with
the evolution of the string network expected to be different from that
of Nambu-Goto strings~\cite{Davis:2000cx}. Calculating the
gravitational radiation from cusps (or cusp-like points) on the chiral
strings, as a function of the current, indicate that the radiation
will be suppressed as compared to nonsuperconducting ones, with the
difference between the gravitational energy radiation from chiral
cusps and nonsuperconducting ones being proportional to the string
current~\cite{Babichev:2003au}. This is expected to lead to important
consequences on the detection of gravitational wave bursts, since a
rather large current seems to result to a lower ampitude of the
incoming signal~\cite{Babichev:2003au}. Stable current-carrying
strings could also be a potential source of primordial magnetic fields
(see e.g. Ref.~\cite{Gwyn:2008fe}).

\subsection{Constraints}
If the strings can support current, it is possible for loops
stabilized by the current, called vortons~\cite{Davis:1988ij}, to
form. These could present a problem as they would prevent decay of the
string network. As noted in Ref.~\cite{Davis:2005jf}, such loops will
be unstable for $m_{\rm F}/m_{\rm D} \lesssim 10^{-2}$, a condition
which is easily satisfied in our case since $m_{\rm F}$ is
exponentially suppressed as compared to $m_{\rm D}$. This means that
the main constraint on the network is the CMB constraint on the string
tension.

The tension of cosmic strings is subject to the bound $G \mu \leq 2
\times 10^{-7}.$ Strings forming at the end of D-term inflation have a
tension $\mu \sim 2 \pi \xi$, where we have argued that $\xi$ must be
given by the Green-Schwarz parameter, i.e.  $ \xi = g^2 \delta_{\rm
  GS}$. Then, in units where $M_{\rm Pl}^2 = 1/(8 \pi G) = 1$, we have
$G \mu = \xi/4$ and the bound is imposed on $g^2 \delta_{\rm GS}$. The
strings satisfy the tension bound only if $g^2 \delta_{\rm GS}<
10^{-7}$. This is the same bound given in
Ref.~\cite{Davis:2005jf}. For the value $\delta_{\rm GS} \sim
10^{-1}$, which is usually quoted~\cite{Dine:1987xk, Atick:1987gy,
  Dine:1987gj}, this is only satisfied in the small coupling regime
which corresponds to a spectral index $n_{\rm s} =1$, larger than the
value currently supported by data.

It is possible that a better understanding of the GS mechanism in the
D3/D7 setup will yield a value of $\delta_{\rm GS}$ substantially smaller
than the usually quoted value; otherwise we are left with the
usual constraints on strings formed at the end of D3/D7 brane inflation.

Possible solutions proposed in the literature have been to make the
strings semilocal, as in Ref.~\cite{Urrestilla:2004eh,
  Dasgupta:2004dw, Chen:2005ae}, since the upper bound on the tension
of semilocal strings is higher than that for local abelian strings, or
to suppress the string production and lower $n_{\rm s}$ by taking
higher order corrections to the K\"ahler potential into account, as in
Ref.~\cite{Haack:2008yb}.

\section{Conclusions}
\label{conclusions}
At present there is no consistent way to include constant FI terms in
a theory coupled to supergravity; only field-dependent FI terms are
allowed. We have explored the implications of this for D3/D7 brane
inflation.  We argue that the D3/D7 setup does make use of a
field-dependent FI term, corresponding to a pseudo-anomalous
U(1)$_{\rm FI}$, so that it is consistent. The main implication of a
pseudo-anomalous U(1)$_{\rm FI}$ for D3/D7 (and indeed D-term)
inflation is that the cosmic strings expected to be formed are the
local axionic strings studied in Ref.~\cite{Davis:2005jf}. In the
D3/D7 case, such strings avoid the problem of vorton production, but
are still subject to the usual CMB bounds. It is possible that further
refinement of the theory could modify the bound somewhat; we leave
this for future work. In the case that current on the strings survives
the gauging of supersymmetry, there may be some interesting
observational effects of such a string network.

\begin{acknowledgments}
It is a pleasure to thank K.~Dasgupta and L.~Uruchurtu for carefully reading the
manuscript. This work is partially supported by the Sciences \&
Technology Facilities Council (STFC--UK), Particle Physics Division,
under the grant ST/G000476/1 ``Branes, Strings and Defects in
Cosmology''. The work of R.~G. is also supported by an NSERC Postdoctoral
Fellowship.
\end{acknowledgments}

\bibliography{FIproject}

\end{document}